\documentclass[aps,preprint,preprintnumber,nofootinbib,amsmath,amssymb,ascmac,12pt]{revtex4}
\usepackage{bm}
\usepackage{graphicx}
\usepackage{color}

\begin{document}
\baselineskip5.5mm
\title{
Evolution of density perturbations in large void universe
}
\author{
       ${}^{1}$Ryusuke Nishikawa \footnote{E-mail:ryusuke@sci.osaka-cu.ac.jp},
       ${}^{2}$Chul-Moon Yoo \footnote{E-mail:yoo@yukawa.kyoto-u.ac.jp},
and
        ${}^{1}$Ken-ichi Nakao \footnote{E-mail:knakao@sci.osaka-cu.ac.jp}
}
\affiliation{
${}^{1}$Department of Mathematics and Physics,
Graduate School of Science, Osaka City University,
3-3-138 Sugimoto, Sumiyoshi, Osaka 558-8585, Japan
\\
${}^{2}$Yukawa Institute for Theoretical Physics,
Kyoto University Kyoto 606-8502, Japan
\vspace{1cm}
}
\begin{abstract}
\baselineskip5.5mm
We study the evolution of linear density perturbations in  
a large spherical void universe which accounts for the acceleration of
the cosmic volume expansion without introducing dark energy. 
The density contrast of this void is not large within the light cone 
of an observer at the center of the void.
Therefore, we describe the void structure as a perturbation with a  
dimensionless small parameter $\kappa$ in a homogeneous 
and isotropic universe within the region observable for the observer. 
We introduce additional anisotropic perturbations 
with a dimensionless small parameter $\epsilon$, whose evolution is of 
interest.
Then, we solve perturbation equations up to order $\kappa \epsilon$
by applying second-order perturbation theory in the homogeneous and 
isotropic universe model. 
By this method, we can know the evolution of anisotropic perturbations 
affected by the void structure. 
We show that the growth rate of the anisotropic density perturbations 
in the large void universe is significantly different from that in the 
homogeneous and isotropic universe. 
This result suggests that the observation of the distribution of galaxies 
may give a strong constraint on the large void universe model.
\end{abstract}

\preprint{OCU-PHYS 363}

\preprint{AP-GR 96}

\preprint{YITP-12-8}

\maketitle

\section{introduction}\label{sec1}

In standard cosmology, the Copernican principle is applied i.e., it is assumed
that we are not living at a privileged position in the universe. 
Combining the Copernican principle with the isotropy of 
the Cosmic Microwave Background (CMB) radiation, 
leads to the conclusion that our universe is well described by the
homogeneous and isotropic universe model. 
In the framework  of the homogeneous and isotropic universe, 
the observational data of the luminosity distances of Type Ia supernovae (SNIa) 
indicates an acceleration of the cosmic volume expansion of our universe.  
The acceleration of the cosmic volume expansion in the homogeneous and 
isotropic universe implies the existence of 
so-called dark energy that acts as a source of a repulsive gravitational force
if we assume general relativity at cosmological scales.
At present, there is no theory that can naturally explain the origin of dark energy, 
and it seems worth investigating alternative scenarios. 
In order to do this, we have to discard general relativity or the homogeneity 
assumption. 

Inhomogeneous cosmological models without dark energy
have been proposed independently 
by Tomita~\cite{Tomita:1999qn,Tomita:2000jj,Tomita:2001gh} 
and C\'el\'erier~\cite{Celerier:1999hp}. 
In C\'el\'erier's model, the observer is located at the symmetry center 
of a very large spherical void which can explain the SNIa observations. 
Since the observer is located at
a special position in the universe, we call this model 
a ``the non-Copernican universe model"  in this paper. 
The common assumption of non-Copernican universe models 
is that an observer is located 
in the vicinity of the symmetry center,
which explains the fact that the CMB radiation is observed to be isotropic. 
The most common way to describe  
non-Copernican universe models 
is to use the Lema\^{\i}tre-Tolman-Bondi(LTB) solution for the Einstein equations, 
which describes the motion of spherically symmetric dust.  

SNIa observations in non-Copernican universe models have been
studied by many researchers~\cite{Tomita:1999qn,Tomita:2000jj,Tomita:2001gh,
Celerier:1999hp,Goodwin:1999ej,Clifton:2008hv,
Iguchi:2001sq,Vanderveld:2006rb,Yoo:2008su,Celerier:2009sv,Kolb:2009hn,Yoo:2010qn}, 
and it has been proven that the distance-redshift relation 
in the $\Lambda$CDM universe can be reproduced using LTB universe 
models~\cite{Mustapha:1998jb,Yoo:2008su,Celerier:2009sv,Kolb:2009hn}. 
Non-Copernican universe models have been tested by other observations including 
the CMB acoustic 
peaks~\cite{Alnes:2005rw,Alexander:2007xx,Zibin:2008vk,
GarciaBellido:2008nz,Yoo:2010qy,Clarkson:2010ej,
Marra:2010pg,Moss:2010jx,Biswas:2010xm,Marra:2011ct,Nadathur:2010zm}, 
the radial baryon acoustic oscillation scales~\cite{Zibin:2008vk,GarciaBellido:2008yq,Zumalacarregui:2012pq},
the kinematic Suniyaev-Zeldovich 
effect~\cite{Bull:2011wi,GarciaBellido:2008gd,Yoo:2010ad,Zhang:2010fa,Moss:2011ze} 
and others~\cite{Alnes:2006pf,Alnes:2006uk,Bolejko:2005fp,Dunsby:2010ts,
Enqvist:2009hn,Enqvist:2006cg,Goto:2011ru,Kodama:2010gr,
Quartin:2009xr,Regis:2010iq,Romano:2009mr,Romano:2010nc,Romano:2011mx,Tanimoto:2009mz,
Uzan:2008qp,Yoo:2010hi,Zibin:2011ma}.
Although these observations have imposed restrictions on these models, 
they have not yet ruled out the models. It is not easy to confirm whether our universe 
follows the Copernican principle.

In this paper, we focus on the evolution of structures such as clusters of galaxies 
and super-clusters in the non-Copernican universe in the matter 
dominant era which is well described by the LTB solution.  
It is expected that observations of the large-scale structures and 
their evolution can be used to test the non-Copernican universe model, 
since the evolution of the anisotropic perturbations reflects the tidal force
in the background spacetime. 
However, the evolution of perturbations in  
the LTB solution has not yet been fully studied. 
This is because the isometries in the LTB spacetime are less than 
in the homogeneous and isotropic universe. 
Although master equations for perturbations for 
general spherically symmetric spacetimes 
have been derived a long time ago~\cite{Gerlach:1980}
(see also Ref.~\cite{Clarkson:2009sc} for the LTB background), 
these equations for the LTB solution cannot be reduced 
to ordinary differential equations, in general.
This is a very different situation from the case of the homogeneous 
and isotropic universe. 

Recently, Alonso et al.~\cite{Alonso:2010zv} performed 
numerical simulations for non-Copernican models including only cold dark matter. 
They studied the perturbed Einstein-deSitter universe with 
two kinds of perturbations: one forms a spherical void, 
and the other is a non-spherical perturbation with a power spectrum with a random phase 
Gaussian probability distribution. 
They followed the growth of these perturbations using Newtonian $N$-body simulations.
However, in order to confirm the validity of the tnumerical simulations, analytic complementary
studies are necessary. 
Some authors~\cite{Moss:2010jx,Zibin:2008vj,Dunsby:2010ts} have studied perturbations
by using a ``silent approximation'' that neglects the magnetic part of the Weyl tensor.
Here, we should note that the magnetic part of the Weyl tensor usually plays an important role 
even in Newtonian situations~\cite{Bertschinger:1994nc}. 
Hence, in this paper, we propose another complementary analytic approach.  

In many non-Copernican models, 
the void structure becomes nonlinear at the present time. 
However, Enqvist et al.~\cite{Enqvist:2009hn} pointed out that 
the void inhomogeneity remains in a quasi-linear regime 
$\sim \mathcal{O}(0.1)$ inside a past light-cone of an observer at the center of the void. 
Actually, they considered a linear perturbation in 
the Einstein-deSitter universe that is consistent with the SNIa data, 
and showed that the fraction of the spherically symmetric 
linear perturbation does not exceed 30\% inside the past light-cone. 
This result implies that non-Copernican LTB cosmological models compatible with 
the observed distance-redshift relation 
may be studied by perturbation theory for the homogeneous and 
isotropic universe filled with dust at least for the inside of 
the past light-cone of the central observer. 

In this paper, we investigate the growth of perturbations in the 
non-Copernican universe models by applying the above idea. 
It is rather difficult to analyze the evolution of anisotropic perturbations 
in the non-Copernican LTB universe model, while it is much easier to study the evolution of 
non-linear perturbations in the homogeneous and isotropic universe model 
by successive approximation. 
We adopt the latter approach. We introduce two-parameter perturbations with 
small expansion parameters $\kappa$ and $\epsilon$
in a homogeneous and isotropic dust universe. 
The limit $\epsilon\rightarrow0$ leads to the exact 
LTB solution, if we take all orders of $\kappa$ into account. By contrast, the limit 
$\kappa\rightarrow0$ with $0<\epsilon\ll1$ leads to the homogeneous and 
isotropic universe with small anisotropic perturbations. 
Then, in order to see the effect of the void structure 
on the evolution of the anisotropic perturbations, we study the non-linear 
effects up to the order of $\kappa\epsilon$, following Ref.~\cite{Tomita:1967}. 

This paper is organized as follows. 
In \S~\ref{sec2}, we derive the equations for 
perturbations parametrized by $\kappa$ and $\epsilon$ 
in the homogeneous and isotropic dust universe and obtain general solutions 
up to order $\kappa\epsilon$. 
In \S~\ref{sec3}, by fixing the initial conditions, 
we calculate the angular power spectrum of the density perturbations. 
In \S~\ref{sec4}, we analyze the growth of the perturbations 
by using the angular growth rate. 
\S~\ref{sec5} is devoted to a summary and discussion.

In this paper, we use the geometrized units in which 
the speed of light and Newton's gravitational constant are one, respectively. 
The Latin indices denote the spatial components, whereas the Greek indices represent
the spacetime components. 
 
\section{two-parameter perturbations in a homogeneous and isotropic dust universe}
\label{sec2}
\subsection{Perturbations with two kinds of parameters}
As mentioned, we study the perturbations in the homogeneous and isotropic universe which is 
often called the Friedmann-Lema\^{\i}tre-Robertson-Walker (FLRW) universe. 
Since the structure formation begins after the universe has begun to be dominated by non-relativistic matter, 
it is sufficient for our purpose to consider the universe model filled with dust. 
Using the spherical polar coordinates for 3-dimensional space, the line-element is given by
\begin{eqnarray}
 d\bar{s}^2 &=& -dt^2+a^2(t)\left(\frac{dr^2}{1-Kr^2}+r^2d\Omega^2\right) =:-dt^2+a^2(t)\gamma_{ij}dx^idx^j,
 \end{eqnarray}
where $a(t)$ is the scale factor which will be determined by the Einstein equations, 
$K$ is constant, $d\Omega^2$ is the 2-dimensional round metric,  
and, for later convenience, we have defined the background conformal 3-metric $\gamma_{ij}$. 
The constant $K$ has the same sign as that of the curvature of the 3-dimensional 
space specified by $t=$constant.
The stress-energy tensor of dust is given by
\begin{equation}
 \bar T^{\mu \nu}= \bar{\rho }(t)\bar{u}^\mu \bar{u}^\nu ,
\end{equation}
where $\bar{\rho}(t)$ is the energy density, 
and $\bar{u}^\mu$ is the 4-velocity whose components are 
given by $\bar{u}^\mu=(1,0,0,0)$. 
The Einstein equations for the FLRW universe are  
\begin{eqnarray}
 \left(\frac{\dot{a}}{a}\right)^2=\frac{8\pi}{3}\bar{\rho }-\frac{K}{a^2} \quad{\rm and}\quad
 \frac{\ddot{a}}{a}=-\frac{4\pi}{3}\bar{\rho},
 \label{bac1}
\end{eqnarray}
where a dot denotes differentiation with respect to $t$. 

Since our main interest is the evolution of density contrasts 
and their correlations, 
we consider only scalar perturbations in the FLRW universe. 
As mentioned in the previous section, 
we introduce two small independent non-negative parameters, $\kappa $ and $\epsilon $. 
The limit $\epsilon\rightarrow0$ leads to the exact 
LTB solution, if we take all the orders of $\kappa$ into account. By contrast, the limit 
$\kappa\rightarrow0$ with $0<\epsilon\ll1$ leads to the homogeneous and 
isotropic universe with small anisotropic perturbations. 

Then, by choosing the synchronous comoving gauge, 
the line element of the perturbed spacetime can be written in the form
\begin{eqnarray}
 ds^2
&=&
 -dt^2+a^2(t)\sum_{N=0}\kappa^N\biggl[
l^{(N)}_{\parallel}(t,r)\frac{dr^2}{1-Kr^2}+l^{(N)}_{\bot }(t,r)r^2d\Omega^2 \cr
&+&\epsilon\Big(A^{(N+1)}(t,r,{\bf \Omega})\gamma _{ij}
+\mathcal{D}_i\mathcal{D}_jB^{(N+1)}(t,r, {\bf \Omega})\Big)dx^idx^j \cr
&+&{\cal O}(\epsilon^2)\biggr],
\label{ds2sec}
\end{eqnarray}
where $l_{\parallel}^{(0)}=l_{\bot}^{(0)}=1$, $\bf \Omega=(\theta,\phi)$ 
are the polar and azimuthal angles, and $\mathcal{D}_i$ denotes the covariant derivative 
with respect to $\gamma_{ij}$. 
The perturbed stress-energy tensor is given by
\begin{equation}
 T^{\mu \nu }
 =
 \bar{\rho }(t)\bar{u}^\mu \bar{u}^\nu 
\sum_{N=0}\kappa^N\left[\Delta ^{(N)}(t,r)+\epsilon \delta ^{(N+1)}(t,r,{\bf \Omega})
+{\cal O}(\epsilon^2)\right],
 \label{tmnsec}
\end{equation}
where $\Delta^{(0)}=1$. 
If we wish to study the evolution of the perturbed FLRW universe 
with the same accuracy as the linear perturbation analysis for the LTB solution, 
we should take all orders of $\kappa$ and the first order with respect to $\epsilon$.  
However, if $\kappa$ is much smaller than unity, 
it will be possible to evaluate the evolution of the anisotropic perturbations in the 
LTB solution by studying up to the first order with respect to $\kappa$.
In this approximation,
the effect of the void structure on the evolution of anisotropic perturbations appears 
at order $\kappa\epsilon$. 

\subsection{First order perturbations}
We expand the perturbation variables in terms of the
spherical harmonic functions $Y_{\ell m}({\bf \Omega})$.
The Einstein equations of order $\kappa$ correspond to the equations for the 
perturbations of $\ell=0$ mode: 
\begin{eqnarray}
 \ddot{\Delta}^{(1)}+2H\dot{\Delta}^{(1)}-4\pi \bar{\rho}\Delta^{(1)} &=&0,
 \label{kap1}
 \\
 \dot{l}^{(1)}_{\parallel}-(r\dot{l}^{(1)}_\bot )^{'}&=&0,
 \label{kap2}
 \\
 \dot{l}^{(1)}_{\parallel}+2\dot{l}^{(1)}_\bot &=&-2\dot{\Delta}^{(1)},
 \label{kap3}
\end{eqnarray}
where 
\begin{equation}
H:=\frac{\dot{a}}{a}, 
\end{equation}
and a dash denotes a partial differentiation with respect to $r$. 
The Einstein equations of order $\epsilon$ lead to the equations for the 
perturbations of $\ell>0$ modes, and we obtain
\begin{eqnarray}
 \dot{A}_{\ell m}^{(1)}(t,r)-K\dot{B}_{\ell m}^{(1)}(t,r)&=&0,
 \label{eps2}
 \\
 (1-Kr^2)\dot{B}_{\ell m}^{(1)}{}''(t,r)
 +\biggl[\frac{2(1-Kr^2)}{r}-Kr\biggr]\dot{B}_{\ell m}^{(1)}{}'(t,r)&& \cr
 +\biggl[3K-\frac{\ell (\ell +1)}{r^2}\biggr]\dot{B}_{\ell m}^{(1)}(t,r) 
 &=&-2\dot{\delta}_{\ell m}^{(1)}(t,r),
 \label{eps3} \\
 \ddot{\delta}_{\ell m}^{(1)}(t,r)+2H\dot{\delta}_{\ell m}^{(1)}(t,r)-4\pi \bar{\rho }\delta_{\ell m}^{(1)}(t,r)&=&0,
 \label{eps1}
\end{eqnarray}
where we have used the eigenvalue equation
\begin{eqnarray}
 & & 
 \left(
 \frac{\partial^2}{\partial\theta ^2}+\frac{\cos \theta }{\sin \theta }\frac{\partial}{\partial\theta}
  +\frac{1}{\sin^2\theta }\frac{\partial^2}{\partial\phi ^2}
 \right)
 Y_{\ell m}({\bf \Omega})=-\ell (\ell +1)Y_{\ell m}({\bf \Omega}).
 \label{eigen}
\end{eqnarray}
We note that the equations of order $\epsilon$ and $\kappa$ 
decouple with each other. 
For later convenience, we also show the equation for $B(t,r,{\bf \Omega})$ before deriving  
Eq.~\eqref{eps3} by the spherical harmonics expansion:
\begin{eqnarray}
 & &
 \big(\mathcal{D}^i\mathcal{D}_i+3K\big)\dot{B}^{(1)}(t,r,{\bf \Omega})
 =-2\dot{\delta}^{(1)}(t,r,{\bf \Omega}). 
 \label{eps4}
\end{eqnarray}

General solutions for Eqs.~\eqref{kap1} and \eqref{eps1} are given by
\begin{eqnarray}
 \Delta ^{(1)}(t,r)
 &=&
 D^+(t)\Delta_+^{(\rm i)}(r)+D^-(t)\Delta_-^{(\rm i)}(r),
 \label{Del1}
 \\
 \delta_{\ell m}^{(1)}(t,r)
 &=&
 D^+(t)\delta^{(\rm i)+}_{\ell m}(r)+D^-(t)\delta^{(\rm i)-}_{\ell m}(r),
 \label{del1}
\end{eqnarray}
where 
\begin{eqnarray}
 & & 
 D^+(t)=H\int^{a(t)}\frac{da}{a^3H^3}\quad{\rm and}\quad D^-(t)=H, 
 \label{grow1}
\end{eqnarray}
and 
$\Delta^{(\rm i)}_{\pm }$ and $\delta^{(\rm i)\pm }_{\ell m}$ stand for initial values.
$D^+$ and $D^-$ represent the growing and decaying modes, respectively.

\subsection{The perturbations of order $\kappa\epsilon$}
As already mentioned, we are interested in 
the effect of the void structure on the evolution of 
anisotropic linear perturbation, and this effect first appears at 
order $\kappa\epsilon$. Hence we shall focus on the perturbations of this order. 

The perturbations of order $\kappa\epsilon$ correspond to the second order 
perturbations in the FLRW universe model. From the second-order Einstein equations together with 
the background and the linearized Einstein equations,
we obtain the evolution equations for the expansion coefficients of $\delta^{(2)}(t,r,{\bf \Omega})$ 
with respect to $Y_{\ell m}({\bf \Omega})$ as follows:
\begin{eqnarray}
 & &
 \ddot{\delta}^{(2)}_{\ell m}(t,r)+2H\dot{\delta}^{(2)}_{\ell m}(t,r)-4\pi \bar{\rho }(t)\delta ^{(2)}_{\ell m}(t,r)
 =
 S_{\ell m}(t,r),
 \label{second1}
\end{eqnarray}
where the source term $S_{\ell m}(t,r)$ is given by
\begin{eqnarray}
 S_{\ell m}(t,r)
 &=&
 \left(\dot{l}^{(1)}_\bot -\dot{l}^{(1)}_{||}\right)\left(K-\frac{\ell (\ell +1)}{2r^2}\right)\dot{B}_{\ell m}^{(1)}
 +\left(\dot{l}^{(1)}_\bot -\dot{l}^{(1)}_{||}\right)\frac{(1-Kr^2)}{r}\dot{B}_{\ell m}^{(1)}{}'
 \nonumber \\
 & &
 +\left(2\dot{\Delta}^{(1)}-\dot{l}^{(1)}_{||}\right)\dot{\delta}^{(1)}_{\ell m}
 +8\pi \bar{\rho}\Delta ^{(1)}\delta ^{(1)}_{\ell m}. 
\end{eqnarray}
By solving Eq.~\eqref{second1}, we obtain 
\begin{eqnarray}
 \delta _{\ell m}^{(2)}(t,r)
 =
\int ^t_{t_{\rm i}}S_{\ell m}(s,r)\left(\frac{D^-(t)D^+(s)-D^+(t)D^-(s)}{W(s)}\right)ds
\end{eqnarray}
where $W(s)$ is the Wronskian given by $W(s)=D^+(s)\dot{D}^-(s)-\dot{D}^+(s)D^-(s)$, 
and homogeneous solutions have been absorbed in $\delta_{\ell m}^{(1)}$. 
Then, we obtain the anisotropic linear density contrast $\delta_{\ell m}$  
in the LTB solution as
\begin{eqnarray}
 \delta_{\ell m}(t,r)
 &=&
 \epsilon \delta^{(1)}_{\ell m}(t,r)
 +\kappa \epsilon
 \int ^t_{t_{\rm i}}S_{\ell m}(s,r)\left(\frac{D^-(t)D^+(s)-D^+(t)D^-(s)}{W(s)}\right)ds \cr
 &+&{\cal O}(\kappa^2\epsilon),
 \label{dens2}
\end{eqnarray}
where $t_{\rm i}$ is the initial time.

Hereafter, we neglect the decaying modes of order $\epsilon$. 
By using Eqs.~\eqref{eps3} and \eqref{del1}, $\dot{B}_{\ell m}^{(1)}$ is 
written in the form 
\begin{eqnarray}
 \dot{B}_{\ell m}^{(1)}(t,r)=\dot{D}^+(t)B^{(\rm i)+}_{\ell m}(r),
 \label{B1}
\end{eqnarray}
where $B^{(\rm i)+}_{\ell m}(r)$ is the initial value.
Then, the anisotropic density contrast \eqref{dens2} is rewritten as
\begin{eqnarray}
 \delta _{\ell m}(t,r)
 &=&
 \epsilon D^+(t)\delta^{(\rm i)+}_{\ell m}(r) \cr
 &+&\kappa \epsilon \left[T_1(t,r)\delta^{(\rm i)+}_{\ell m}(r)
 +T_2(t,r,\ell )B^{(\rm i)+}_{\ell m}(r)
 + T_3(t,r)B^{(\rm i)+}_{\ell m}{}'(r)
 \right] \cr
 &+&{\cal O}(\kappa^2\epsilon),
 \label{dens3}
\end{eqnarray}
where $T_1$, $T_2$ and $T_3$ are defined by
\begin{eqnarray}
 T_1(t,r)
 &:=&
 \int ^t_{t_{\rm i}}ds\left(\frac{D^-(t)D^+(s)-D^+(t)D^-(s)}{W(s)}\right)
 \nonumber \\
 & &
 \times
 \left[\dot{D}^+(s)\left(2\dot{\Delta}^{(1)}(s,r)-\dot{l}^{(1)}_{||}(s,r)\right)
 +D^+(s)\times 8\pi \bar{\rho}(s)\Delta ^{(1)}(s,r)
 \right],
 \\
 T_2(t,r,\ell )
 &:=&
 \int ^t_{t_{\rm i}}ds\left(\frac{D^-(t)D^+(s)-D^+(t)D^-(s)}{W(s)}\right)
 \nonumber \\
 & &
 \times
 \dot{D}^+(s)
 \left(\dot{l}^{(1)}_\bot(s,r) -\dot{l}^{(1)}_{||}(s,r)\right)\left(K-\frac{\ell (\ell +1)}{2r^2}\right),
 \\
 T_3(t,r)
 &:=&
 \int ^t_{t_{\rm i}}ds\left(\frac{D^-(t)D^+(s)-D^+(t)D^-(s)}{W(s)}\right)
 \nonumber \\
 & &
 \times
 \dot{D}^+(s)\left(\dot{l}^{(1)}_\bot (s,r)-\dot{l}^{(1)}_{||}(s,r)\right)\frac{1-Kr^2}{r}.
\end{eqnarray}

\section{Angular power spectrum and angular growth rate}\label{sec3}
In the previous section, we derived the growing solutions for density contrasts. 
Once we have specified the isotropic linear perturbations 
$l^{(1)}_{||}$, $l^{(1)}_{\bot }$, $\Delta ^{(1)}$ and 
the initial anisotropic inhomogeneities 
$\delta^{(\rm i)+}_{\ell m}$, $B^{(\rm i)+}_{\ell m}$, 
we obtain the density contrasts $\delta_{\ell m}$ by Eq.~\eqref{dens3}. 
In this section, 
we derive the explicit form of the angular power spectrum of 
the density perturbation $\delta_{\ell m}$ 
for a given set of initial conditions in terms of the standard power spectrum. 
Then, we define the angular growth rate by using the angular power spectrum. 
Hereafter, in order to determine the perturbations of order $\kappa$,
we refer to the non-Copernican LTB universe model with uniform big-bang time (see Appendix A).

\subsection{Initial power spectrum of density contrast} 
By virtue of the uniform big-bang time (see Appendix A),
the present non-Copernican LTB universe 
approaches the homogeneous and isotropic universe as time goes back.  
Hence, it is reasonable to assume that the initial conditions for the anisotropic 
perturbations are the same as in the case of the FLRW universe.
Then, the initial power spectrum of the density contrast can be 
expressed as follows: 
\begin{eqnarray}
 \langle\delta ^{(1)*}(t_{\rm i},{\bf k})\delta ^{(1)}(t_{\rm i},{\bf k}')\rangle
 =(2\pi)^3\delta_{\rm D}^3({\bf k}-{\bf k}')P(t_{\rm i},k),
 \label{pow1}
\end{eqnarray}
where $\delta_{\rm D}$ is the Dirac's delta function, 
$t_{\rm i}$ represents some sufficiently early time already introduced in Eq.~\eqref{dens2},
and the Fourier transform of the density contrast is defined by
\begin{eqnarray}
 \delta^{(1)} (t,{\bf k})=\int d^3x \delta^{(1)} (t,{\bf x})e^{-i{\bf k\cdot x}}.
\end{eqnarray}
If we choose the initial time after recombination, 
the matter power spectrum including baryons and cold dark matter can be written as
\begin{eqnarray}
 P(t_{\rm i},k) &=& [D^+(t_{\rm i})]^2P(k), \nonumber \\
 P(k)&=&A_0k^nT^2(k),
 \label{pow2}
\end{eqnarray}
where $A_0$ is a positive constant which represents the amplitude for perturbations on large scales, 
$n$ is constant, and $T(k)$ is the matter transfer function.
In this paper, we assume the Harrison-Zel'dovich spectrum $n=1$. 
As for the transfer function,
we adopt the fitting formula developed by Eisenstein \& Hu~\cite{Eisenstein:1997ik} 
(see Appendix B). 

\subsection{Angular power spectrum and angular growth rate}
In order to observationally study the evolution of perturbations in the non-Copernican universe,
we need to specify the observable quantities by using the density contrast \eqref{dens3}.
The simplest quantity that we can currently calculate is the angular power spectrum.
We define the angular power spectrum of the density contrast by 
\begin{eqnarray}
 C_\ell (t,r)=
 \frac{r^2}{2\ell +1}\sum_{m=-\ell}^\ell \langle\delta^{*}_{\ell m}(t,r)\delta_{\ell m}(t,r)\rangle,
 \label{CL1}
\end{eqnarray}
where $*$ denotes the complex conjugate.

Hereafter, we focus on the non-spherical perturbations whose wavelengths 
are much smaller than the spatial curvature radius $(k\gg \sqrt{|K|})$. 
Then, from Eq. \eqref{eps4}, we have
\begin{eqnarray}
 k^2\dot{B}^{(1)}(t,{\bf k})\simeq 2\dot{\delta}^{(1)}(t,{\bf k}). 
 \label{fou}
\end{eqnarray}
The initial values $\delta^{(\rm i)}_{\ell m}(r)$ and $B^{(\rm i)}_{\ell m}(r)$ 
which appear in Eqs.~\eqref{B1} and \eqref{dens3} 
can be written using the Fourier transform of the initial density contrast $\delta^{(1)}(t_{\rm i},{\bf k})$ as
\begin{eqnarray}
 \delta^{(\rm i)+}_{\ell m}(r)
 &=&
 \Big[\frac{1}{D^+(t_{\rm i})}\Big](4\pi i^\ell )\int \frac{d^3k}{(2\pi)^3}\delta^{(1)} (t_{\rm i},{\bf k})j_\ell (kr)Y^*_{\ell m}
 ({\bf \Omega}_{\rm k}),
 \label{de2}
 \\
 B^{(\rm i)+}_{\ell m}(r)
 &=&
 \Big[\frac{1}{D^+(t_{\rm i})}\Big](4\pi i^\ell )\int \frac{d^3k}{(2\pi)^3}\delta^{(1)} (t_{\rm i},{\bf k})j_\ell (kr)
 Y^*_{\ell m}({\bf \Omega}_{\rm k})\left(\frac{2}{k^2}\right),
 \label{B2}
\end{eqnarray}
where ${\bf \Omega}_{\rm k}$ denotes the polar and azimuthal angles in the Fourier space, and 
we have used the relation between the expansion coefficient with respect to  
$Y_{\ell m}({\bf \Omega}_{\rm k})$ and the Fourier transform
\begin{eqnarray}
 \phi_{\ell m}(t,r)
 =
 (4\pi i^\ell )\int \frac{d^3k}{(2\pi)^3}\phi (t,{\bf k})j_\ell (kr)Y^*_{\ell m}({\bf \Omega}_{\rm k}),
 \label{phi2}
\end{eqnarray}
and Eq.~\eqref{fou}. 
Using Eqs.~\eqref{pow1}, \eqref{pow2}, \eqref{de2} and \eqref{B2}, 
the angular power spectrum of the density contrast \eqref{dens3} 
can be rewritten in the following form: 
\begin{eqnarray}
C_\ell (t,r)&=&
\epsilon^2{D^+}^2(t)K_1(\ell ,r)+\kappa \epsilon^2 \Big[2D^+(t)T_1(t,r)K_1(\ell ,r)
\nonumber \\
& &
+2D^+(t)\tilde{T}_{2}(t,r,\ell )K_2(\ell ,r)+2D^+(t)\tilde{T}_{3}(t,r)K_3(\ell ,r)
\Big]+{\cal O}(\kappa^2\epsilon^2),
\label{CL4}
\end{eqnarray}
where 
\begin{eqnarray}
 K_1(\ell ,r)
 &=&
 \left(\frac{2}{\pi }\right)\int _0^\infty dk P(k)(kr)^2j_\ell ^2(kr),
 \nonumber \\
 K_2(\ell ,r)
 &=&
 \left(\frac{2}{\pi }\right)\int _0^\infty dk P(k)(kr)^2j_\ell ^2(kr)\left(\frac{2}{k^2}\right),
 \nonumber \\
 K_3(\ell ,r)
 &=&
 \left(\frac{2}{\pi }\right)\int _0^\infty dk P(k)(kr)^2j_\ell (kr)j_\ell ^{'}(kr)\left(\frac{2}{k^2}\right),
\end{eqnarray}
and $\tilde{T}_2$ and $\tilde{T}_3$ are defined 
as the short wavelength approximation $(k\gg \sqrt{|K|})$ of 
$T_2$ and $T_3$ by 
\begin{eqnarray}
 \tilde{T}_2(t,r,\ell )
 &=&
 \int ^t_{t_{\rm i}}ds\left(\frac{D^-(t)D^+(s)-D^+(t)D^-(s)}{W(s)}\right)
 \nonumber \\
 & &
 \times
 \dot{D}^+(s)
 \left(\dot{l}^{(1)}_\bot(s,r) -\dot{l}^{(1)}_{\parallel}(s,r)\right)\left(-\frac{\ell (\ell +1)}{2r^2}\right),
 \\
 \tilde{T}_3(t,r)
 &=&
 \int ^t_{t_{\rm i}}ds\left(\frac{D^-(t)D^+(s)-D^+(t)D^-(s)}{W(s)}\right)
 \nonumber \\
 & &
 \times
 \dot{D}^+(s)\left(\dot{l}^{(1)}_\bot (s,r)-\dot{l}^{(1)}_{||}(s,r)\right)\left(\frac{1}{r}\right).
\end{eqnarray} 
Once the initial density power spectrum $P(t_{\rm i},k)$ is specified, 
we can calculate the angular power spectrum $C_\ell (t,r)$ by using Eq.~\eqref{CL4}.

To investigate the growth rates of the perturbations, 
we define the angular growing factor by 
\begin{eqnarray}
 D_\ell(t,r)=\left[\frac{C_\ell(t,r)}{C_\ell(t_{\rm i},r)}\right]^{1/2}. 
 \label{eff1}
\end{eqnarray}
It is easy to see that the angular growing factor $D_\ell(t,r)$ is equal to $D^+(t)/D^+(t_{\rm i})$ 
up to order $\epsilon$. 

Basically, an observer can see the cosmological structures on his/her past light cone 
by observations through electromagnetic 
radiation\footnote{The past light cone of an observer at the event $p$ 
is defined by the boundary of the causal past of $p$, which is usually 
denoted by $\dot{J}^-(p)$ in general relativity. 
Strictly speaking, the observer can see the inside of the light cone 
through a congruence of the light rays which have experienced caustics caused by gravitational 
lens effects or scattering due to electromagnetic interactions in the real universe.}
and hence, in the case of the non-Copernican universe model, 
it is useful to consider quantities on the light cone of an observer who stays 
at the symmetry center of the void at present. 
Hereafter, for simplicity, we call the observer who stays at the symmetry center of 
the void at present ``the central observer", and the past light cone of the central observer 
is denoted by $\Sigma_{\rm lc}$. 
The past light cone $\Sigma_{\rm lc}$ is generated by the past-directed outgoing 
radial null geodesics $k^\mu=(dt/d\lambda,dr/d\lambda,0,0)$, where $\lambda$ is 
the affine parameter. The cosmological redshift $z$ is defined by
\begin{equation}
z=\frac{dt/d\lambda}{(dt/d\lambda)_0}-1,
\end{equation}
where $dt/d\lambda$ and $(dt/d\lambda)_0$ are the value at the time of the emission of 
a photon and that at the time of the detection of the photon by the central observer, respectively. 
By using the cosmological redshift $z$ instead of 
the affine parameter $\lambda$, the geodesic equations for the generator of the 
past light cone $\Sigma_{\rm lc}$ up to order $\kappa$ are given by
\begin{eqnarray}
\frac{dr}{dz}&=&\frac{\sqrt{1-Kr^2}}{(1+z)aH}\left[1-\frac{\kappa}{2}
\left(l_\parallel+\frac{1}{H}\dot{l}_\parallel\right)\right], \\
\frac{dt}{dz}&=&-\frac{1}{(1+z)H}\left(1-\frac{\kappa}{2H}\dot{l}_\parallel\right). 
\end{eqnarray}
We denote the solution of the above equations by $t=t_{\rm lc}(z)$ and $r=r_{\rm lc}(z)$.  

Then, using the angular growing factor, we define the 
angular growth rate on the past light cone $\Sigma_{\rm lc}$ as 
a function of redshift $z$ as follows:
\begin{eqnarray}
 f_\ell(z)=- \frac{d [\ln D_\ell(t_{\rm lc}(z),r_{\rm lc}(z))]}{d \ln(1+z)}. 
 \label{eff2}
\end{eqnarray}
Here, we note that the angular growing factor and the angular growth rate 
do not depend on the amplitude $A_0$ in Eq.~\eqref{pow2}. 
We also note that the angular growth rate up to order $\epsilon$ agrees with the growth rate 
usually used in the linear perturbation theory of the FLRW universe, 
 $d ({\rm ln} D^+)/d ({\rm ln} a)$.

\section{evolution of density perturbations in the Clarkson-Regis model}\label{sec4}
\subsection{Linearized Clarkson-Regis model}
In order to determine the perturbations of order $\kappa$, 
we use the non-Copernican LTB universe model 
given by Clarkson and Regis~\cite{Clarkson:2010ej} 
(see Appendix A), which we call the Clarkson and Regis~(CR) model. 
We shall study the evolution of linear anisotropic perturbations 
in the CR model by using the second order perturbation theory of the FLRW universe filled with dust. 
In order to approximate the CR model by the linearly perturbed FLRW 
universe filled with dust, we must first specify the background FLRW universe.  
Here, we determine the background FLRW universe so that 
the cosmological density parameter of the background is equal to 
$0.242$, which is equal to the value of the density parameter function $\Omega_{\rm M}(r)$
at the symmetry center of  the CR model (see Eq.~\eqref{CR1}).

\begin{figure}[htbp]
 \begin{center}
 \includegraphics[width=10cm,clip]{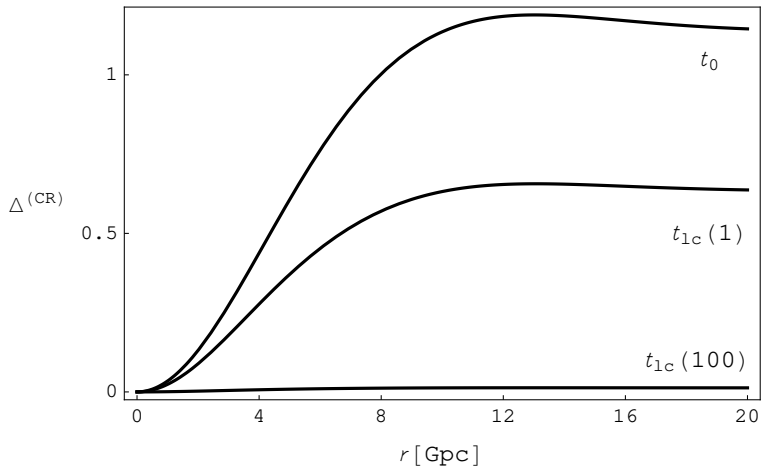}
 \end{center}
 \caption{
Density contrasts on the spacelike hypersurfaces for
$t=t_{\rm lc}(100)$, $t=t_{\rm lc}(1)$ 
and $t=t_0$, as functions of $r$.
}
 \label{density_timeconst}
\end{figure}

We define the ``density contrast'' of the CR model as
\begin{eqnarray}
 & &
 \Delta^{(\rm CR)}(t,r)= \frac{\rho^{(\rm CR)}(t,r)-\rho^{\rm (CR)}(t,0)}{\rho^{\rm (CR)}(t,0)},
 \label{Del2}
\end{eqnarray}
where $\rho^{(\rm CR)}$ is the energy density of 
the CR model. The density contrasts $\Delta^{(\rm CR)}$'s on three constant time 
hypersurfaces are depicted in Fig.~\ref{density_timeconst} as functions of $r$. 
Here, we have used the cosmological redshift $z$ to specify each constant 
time hypersurface given by $t=t_{\rm lc}(z)$.
We can see that the void structure grows with time. 
Since the big-bang time is uniform, there is only the growing mode in the CR model. 
The void size is about $12{\rm Gpc}$, and the vicinity of the center 
is locally the dust filled FLRW model 
with the cosmological density parameter $\Omega_{\rm M}=0.242$, whereas 
the asymptotic region is almost the same as the dust filled FLRW model with $\Omega_{\rm M}=0.7$.
The Hubble parameter at the center is $H_{0}=74{\rm kms^{-1}Mpc^{-1}}$. 

By using the density contrast $\Delta^{(\rm CR)}$, 
we give the initial conditions for the isotropic linear density contrast 
$\Delta ^{(\rm i)}_{\pm }$ in Eq.~\eqref{Del1} as follows.
As mentioned, since the CR model has only the growing mode,  
we should set $\Delta ^{(\rm i)}_{-}(r)=0$. By contrast, 
$\Delta_+^{({\rm i})}(r)$ is determined by the assumption that  
the density contrast $\Delta^{(1)}$ exactly agrees 
with that of the CR model at the initial time $\Delta^{\rm (CR)}(t_{\rm i},r)$, i.e.,
\begin{eqnarray}
 \Delta ^{(\rm i)}_+(r)=\frac{\Delta^{(\rm CR)}(t_{\rm i},r)}{D^+(t_{\rm i})},
\end{eqnarray}
where the initial time is determined by $t_{\rm i}=t_{\rm lc}(1000)$. 
Once the initial condition for the density contrast is given, 
we obtain all perturbations of order $\kappa$ 
by solving the perturbation equations up to the corresponding order, and as a result, 
the linearized CR model is obtained. 

\begin{figure}[htbp]
 \begin{center}
 \includegraphics[width=10cm,clip]{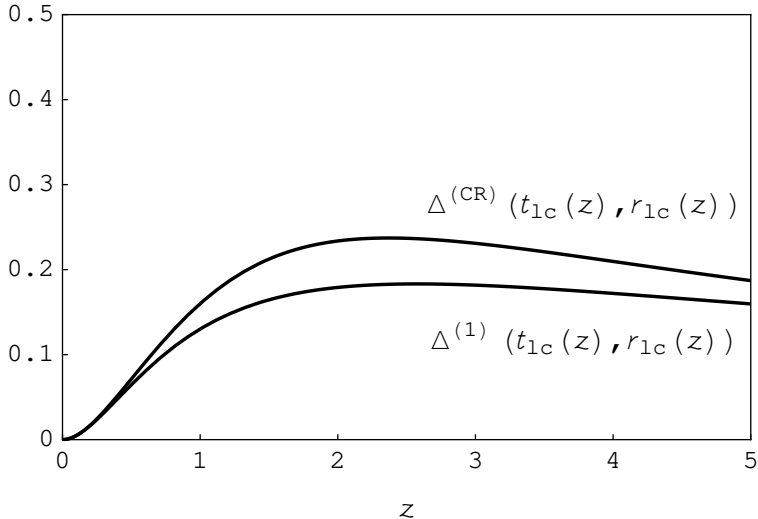}
 \end{center}
 \caption{
 Exact density contrast $\Delta^{(\rm CR)}$ and 
 that of the linearized CR model $\Delta^{(1)}$ on the past light cone $\Sigma_{\rm lc}$,
 plotted as functions of the redshift $z$.
 }
 \label{density_redshift}
\end{figure}

To evaluate the accuracy of the linear approximation,
we plot $\Delta^{(\rm CR)}$ and $\Delta^{(1)}$ on the past light cone $\Sigma_{\rm lc}$ 
in Fig.~\ref{density_redshift}.
The relative error between the exact and linearized CR models is less than $30\%$
on the light cone $\Sigma_{\rm lc}$. 
Inside the past light cone $\Sigma_{\rm lc}$, 
the error is smaller than that on the past light cone $\Sigma_{\rm lc}$,
since the CR model has only the growing mode. 
There is no qualitative difference between the exact and the linearized CR models,
and thus we may see the qualitative behavior of linear perturbations in the CR model by  
the perturbative analysis of the FLRW universe based on the linearized CR model. 

\subsection{Evolution of density contrasts in the CR model}
\begin{figure}[htbp] 
 \begin{center}
\includegraphics[width=10cm,clip]{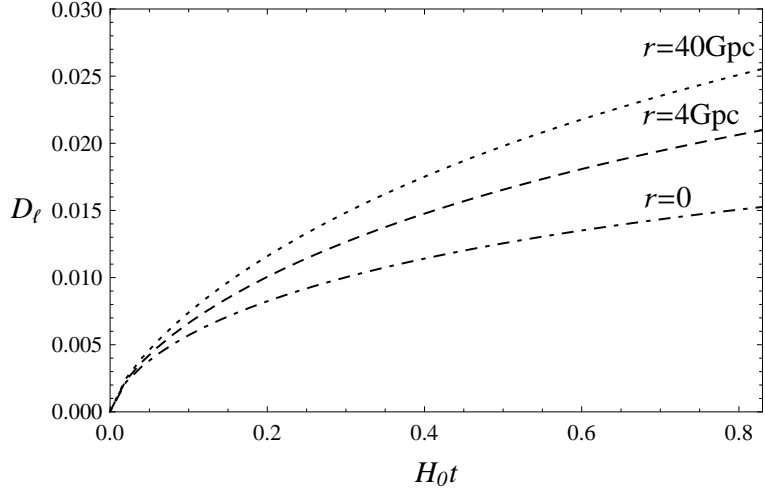}
 \end{center}
 \caption{
Angular growing factors $D_\ell$'s in the CR model at 
$r=40{\rm Gpc}$ (dotted line), $r=4{\rm Gpc}$ (dashed line) and $r=0$ (dot-dashed line)
depicted as functions of $t$. 
The present time is $H_0t_0=0.83$. 
We choose $\ell$ so that $\tilde{k}=0.5 {\rm Mpc}^{-1}$.
}
 \label{growthfactor_CR}
\end{figure}

Let us consider the evolution of the anisotropic density contrasts in the CR model. 
By using the angular power spectrum $C_\ell(t,r)$ given by Eq.~\eqref{CL4} and 
the transfer function $T(k)$ given in appendix~\ref{appendixB},
we depict the angular growing factors $D_\ell(t,r)$'s defined by Eq.~\eqref{eff1} 
at each comoving distance as functions of $t$ in Fig.~\ref{growthfactor_CR}. 

\begin{figure}[htbp]
 \begin{center}
 \includegraphics[width=10cm,clip]{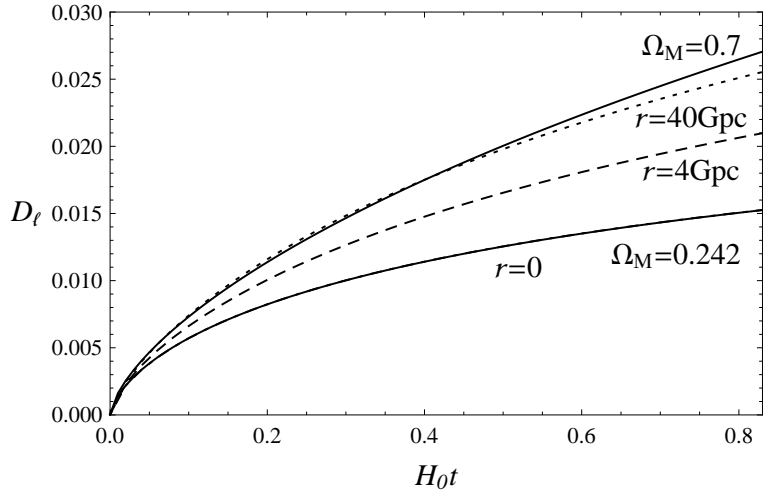}
 \end{center}
 \caption{
Angular growing factors $D_\ell$'s in the dust filled FLRW universe models 
with $\Omega_{\rm M}=0.242$ and $\Omega_{\rm M}=0.7$, together with that  
for the CR model. 
}
 \label{growthfactor_FLRW}
\end{figure}

Here, we introduce a useful quantity defined by  
\begin{equation}
{\tilde k}:=\frac{\ell}{r}~.
\end{equation} 
Note that $\tilde k$ is equal to the comoving wave number of the mode 
$\ell$ at a distance $r$ in the flat sky approximation 
(see e.g., Ref.~\cite{Lyth:textbook}).
In Fig.~\ref{growthfactor_CR}, the present time is $H_0t_0=0.83$, and $\ell$ is 
chosen so that $\tilde{k}=0.5 {\rm Mpc}^{-1}$.  
This choice of $\ell$ shows us the evolution of perturbations with the size of a cluster of galaxies, i.e.,  
$2\pi /\tilde{k}  \sim 10 {\rm Mpc}$.

We can see from Fig.~\ref{growthfactor_CR} that 
the larger the comoving distance of a perturbation from the symmetry center, 
the faster the growth of the perturbation. 
This result may be explained by the fact that the energy density of the CR model 
is a monotonically increasing function of $r$, since the growth rates of 
perturbations in the FLRW universe is a monotonically increasing function of $\Omega_{\rm M}$. 
We also depict $D_\ell$ of the FLRW universe models   
with $\Omega_{\rm M}=0.242$ and $0.7$, respectively,  
together with that of the CR model in Fig.~\ref{growthfactor_FLRW}. 
We can see from this figure that $D_\ell$ of the FLRW universe with 
$\Omega_{\rm M}=0.242$ agrees with  $D_\ell(t,r=0)$ of the CR model. 
We note that $D_\ell$ in the FLRW universe with $\Omega_{\rm M}=0.7$ 
does not agree with that far from the void ($r=40{\rm Gpc}$) in the CR  model.
This result might not be real, but rather could be an error caused by using 
the linearized CR model, since the error due to the linear
approximation becomes larger for larger $r$.  

\begin{figure}[htbp]
 \begin{center}
 \includegraphics[trim = 5 110 5 110 ,width=14cm,clip]{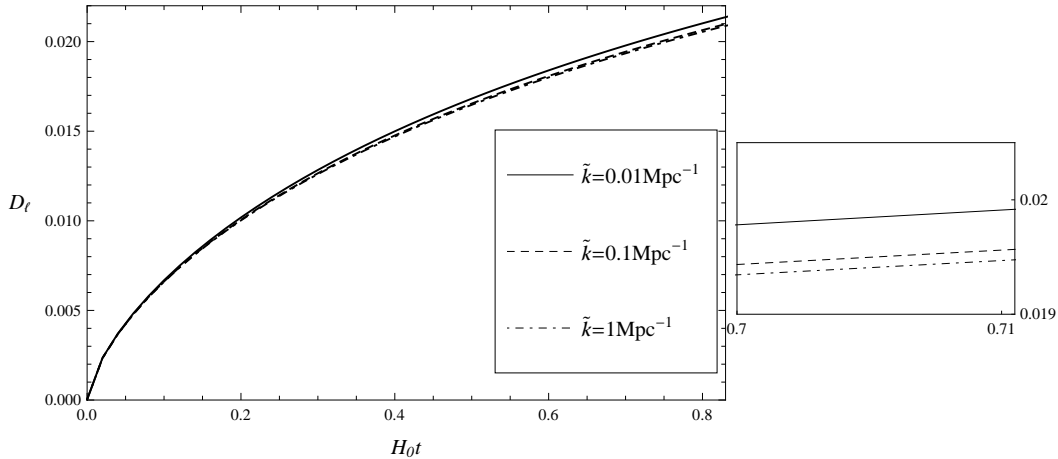}
 \end{center}
 \caption{
  Angular growing factors $D_\ell(t,r=4{\rm Gpc})$ in the CR model for
  $\tilde{k}=0.01{\rm Mpc^{-1}}$ (solid line), $0.1{\rm Mpc^{-1}}$ (dashed line) 
  and $1{\rm Mpc^{-1}}$ (dot-dashed line), depicted 
  as functions of $t$.
  The right panel shows a close-up of the $0.7<H_0t<0.71$}.
 \label{kdepend}
\end{figure}

We have also investigated the dependence of $D_\ell$ 
on $\tilde{k}$. 
The angular growing factors $D_\ell(t,r=4{\rm Gpc})$'s 
with various values of $\ell$, or equivalently, 
$\tilde{k}$ are depicted as functions of $t$ in Fig.~\ref{kdepend}.
We find that the dependence of $D_\ell(t,r=4{\rm Gpc})$ on 
$\tilde{k}$ is very small in the case of the CR model.

\begin{figure}[htbp]
 \begin{center}
 \includegraphics[width=10cm,clip]{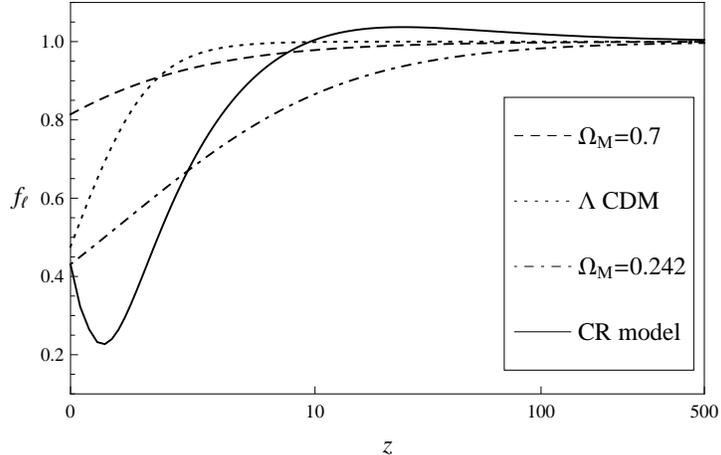}
 \end{center}
 \caption{Angular growth rate $f_\ell$ in the CR model (solid line), 
together with those of 
the dust filled FLRW universe with $\Omega_{\rm M}=0.242$ (dot-dashed line) and 
$\Omega_{\rm M}=0.7$ (dashed line)
and the flat $\rm {\Lambda CDM}$ universe with $\Omega_{\rm M}=0.28$ (dotted line)
as a function of $z$. $\ell$ is chosen so that $\tilde{k}=0.5 {\rm Mpc}^{-1}$.
}
 \label{growthrate_high}
\end{figure}

\begin{figure}[htbp]
 \begin{center}
 \includegraphics[width=10cm,clip]{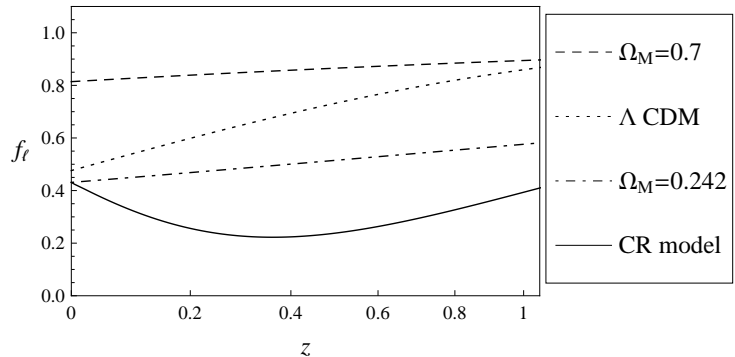}
 \end{center}
 \caption{
Close-up of the region $0<z<1$ in Fig.~\ref{growthrate_high}.
}
 \label{growthrate_low}
\end{figure}

Next, the angular growth rate $f_\ell$ defined by \eqref{eff2} is plotted as 
a function of $z$ for the CR model, together with those of  
the dust filled FLRW with $\Omega_{\rm M}=0.242$ and $\Omega_{\rm M}=0.7$ 
and the flat $\rm {\Lambda CDM}$ with $\Omega_{\rm M}=0.28$
in Figs.~$\ref{growthrate_high}$ and $\ref{growthrate_low}$.
Figure~\ref{growthrate_low} shows a close-up of the region $0<z<1$ in 
Fig.~\ref{growthrate_high}.  
Here, $\ell$ is also chosen so that $\tilde{k}=0.5 {\rm Mpc}^{-1}$.
From Fig.~\ref{growthrate_high},
we can see that the value of $f_\ell$ of these models approach
the value of the Einstein de-Sitter universe ($f_\ell=1$) in the high redshift domain. 
We can see from Fig.~\ref{growthrate_low} that $f_\ell$ at the central observer $z=0$ of the CR model 
agrees with the value of the dust filled FLRW universe with $\Omega_{\rm M}=0.242$. 
We can also see that $f_\ell$ in the CR model 
is significantly different from those in homogeneous and 
isotropic universes for redshift $0<z\lesssim 1$.
This result implies that, if we can somehow observe $f_\ell$,  
the observational data may give a strong constraint on non-Copernican universe models.

\section{summary and discussion}\label{sec5}
We have studied the evolution of anisotropic density perturbations 
in a large spherical void model, 
which is a non-Copernican universe model based on the LTB solution.  
As is well known, the analysis of anisotropic linear perturbations of the LTB solution 
is much harder than that of the homogeneous and isotropic universe. 
Therefore, we have not tried to directly solve linear perturbations in the LTB model. 
Instead, we have studied linear perturbations in a linearized LTB model.
The linearized LTB model is obtained by regarding the inhomogeneities 
in the LTB model as isotropic linear perturbations in the dust filled FLRW universe.  
Hence, our method is relevant for the LTB model only if the isotropic inhomogeneities is so small 
that they can be treated as linear perturbations on a dust filled FLRW universe.

Assuming the uniform big-bang time, 
we have given the initial conditions for anisotropic perturbations, which are the same as 
those of the FLRW universe model. 
Then, we calculated the angular power spectrum of the anisotropic density contrast
taking into account the tidal effect of isotropic inhomogeneity, 
and further we defined the angular growing factor and an angular growth rate 
using the angular power spectrum. 

In \S~\ref{sec4}, 
we calculated the angular growing factor using the
linearized Clarkson-Regis (CR) model, which has the uniform big-bang time.
From the behavior of the angular growing factor,
we found that the speed of growth of a perturbation is a monotonically increasing 
function of the comoving distance from the center of the void. 
Because of this property, the angular growth rate in the CR model differs from 
that in the dust filled FLRW universe even for low redshift $(z<1)$.
So, if we can observe the angular power spectrum of 
the density contrasts and can observe their angular growth rate,
the observational data will strongly restrict non-Copernican universe models.

Guzzo et al. and Blake et al. have presented the observational results on the 
growth rate of the density perturbation as a function of the redshift $z$ 
on our past light cone~\cite{Guzzo:2008ac,Blake:2011rj}. 
Here, we should note that their observations are a distortion of the correlation function 
of galaxies in redshift space and they derived the growth rate from it 
assuming the homogeneous and isotropic background universe. 
Hence, our present result regarding the angular growth rate does not correspond to 
their ``observational results on the growth rate''. 
To calculate the redshift space distortion of the galaxy distribution 
in non-Copernican universe models,
we have to consider not only the growth in the transverse 
direction but also the radial direction. 
This is a future work, and we shall present the results elsewhere. 

We expect that the formula developed here 
could be applied to many inhomogeneous cosmological models,
since we have only assumed that the isotropic 
inhomogeneities are small inside the past light cone $\Sigma_{\rm lc}$. 
We also expect that the formula could be potentially used for 
calculating the integrated Sachs-Wolfe effect, 
baryon acoustic oscillations, the shape of the 3D power spectrum and so on. 

\section*{Acknowledgments}

We are grateful to H.~Ishihara and colleagues in the astrophysics and 
gravity group of Osaka City University for their useful and helpful 
discussion and criticism. 
RN and CY are supported by a Grant-in-Aid through the
Japan Society for the Promotion of Science (JSPS).
This work was partially supported by a JSPS Grant-in-aid for Scientific Research (C) (No.\ 21540276).

\appendix
\allowdisplaybreaks
\section{Clarkson-Regis model}\label{appendixA}

The Clarkson-Regis (CR) model is a non-Copernican universe model based on the 
Lema\^{\i}tre-Tolman-Bondi (LTB) spacetimes, which is an exact solution of the 
Einstein equations and describes the motion of spherically symmetric dust. 
The CR model can explain the SNIa data and peak positions of 
the fluctuations in the CMB radiation. 
The line element and the stress-energy tensor of the LTB spacetime are given by
\begin{eqnarray}
 ds^2
 &=&
 -dt^2+\frac{(\partial_rR(t,r))^2}{1-k(r)r^2}dr^2+R^2(t,r)d\Omega^2,
 \label{LTB:ds}
 \\
 T^{\mu \nu}
 &=&
 \rho(t,r)u^\mu u^\nu ,
 \label{LTB:tmn}
\end{eqnarray}
where 
$k$, $u^\mu$ and $\rho$ are 
an arbitrary function of the radial coordinate $r$, 
the 4-velocity and the energy density of the dust, respectively. 
The 4-velocity of the dust fluid is given by $u^\mu =\delta^\mu_0$.
The Einstein equations lead to
\begin{eqnarray}
 \left(\frac{\partial_tR}{R}\right)^2
 &=& \frac{2M(r)}{R^3}-\frac{k(r)r^2}{R^2},
 \label{LTB1} \\
 \rho(t,r)
 &=&
 \frac{\partial_rM(r)}{4\pi R^2(t,r)\partial_rR(t,r)},
\end{eqnarray}
where $M(r)$ is an arbitrary function of $r$.
By integrating Eq.~\eqref{LTB1}, 
we obtain 
\begin{eqnarray}
 R(t,r)=(6M(r))^{1/3}(t-t_{\rm B}(r))^{2/3}S(x),
 \end{eqnarray}
where $t_{\rm B}(r)$ is an arbitrary function of $r$, $x$ is defined by
\begin{eqnarray}
 x
: =
 k(r)r^2\left(\frac{t-t_{\rm B}(r)}{6M(r)} \right)^{2/3},
\end{eqnarray}
and, by defining $\eta$ as
\begin{equation}
x=:
  \begin{cases}
  \dfrac{-(\sinh\sqrt{-\eta}-\sqrt{-\eta})^{2/3}}{6^{2/3}}
   & \text{for~~~} x<0, \rule{0pt}{20pt} \\
   \dfrac{(\sqrt{\eta}-\sin\sqrt{\eta})^{2/3}}{6^{2/3}}
   & \text{for~~~} x>0, \rule{0pt}{26pt} 
  \end{cases}
\end{equation}
the function $S(x)$ is given by
\begin{equation}
 S(x)=
  \begin{cases}
   \dfrac{\cosh\sqrt{-\eta }-1}{6^{1/3}(\sinh\sqrt{-\eta}-\sqrt{-\eta})^{2/3}} 
   & \text{for~~~} x<0, \rule{0pt}{20pt} \\
   \dfrac{1-\cos\sqrt{\eta }}{6^{1/3}(\sqrt{\eta}-\sinh\sqrt{\eta})^{2/3}}
   & \text{for~~~} x>0, \rule{0pt}{26pt} 
  \end{cases}
\end{equation}
and $S(0)=(3/4)^{1/3}$.

The radial and azimuthal Hubble rates are defined by
\begin{eqnarray}
 H_\parallel(t,r):=\frac{\dot{R}'(t,r)}{R'(t,r)}
 \quad {\rm and} \quad
 H_\bot (t,r):=\frac{\dot{R}(t,r)}{R(t,r)}.
\end{eqnarray}
By defining the density-parameter function as
\begin{eqnarray}
 \Omega_{\rm M}(r):=\frac{2M(r)}{H_{\bot 0}^2(r)R_0^3(r)},
 \label{omg}
\end{eqnarray}
we can rewrite Eq.~\eqref{LTB1} in a form similar to the Friedmann equation:
\begin{eqnarray}
 H_\bot ^2(t,r)=H_{\bot 0}^2(r)\Big[\Omega_{\rm M}(r)\Big(\frac{R_0(r)}{R(t,r)}\Big)^3+(1-\Omega_{\rm M}(r))\Big(\frac{R_0(r)}{R(t,r)}\Big)^2\Big],
 \label{LTB2}
\end{eqnarray}
where functions with subscripts 0 correspond to present values, $H_{\bot 0}(r)=H_\bot (t_0,r)$ and $R_0(r)=R(t_0,r)$.
By integrating \eqref{LTB2}, we have
\begin{eqnarray}
 H_{\bot 0}(r)=\frac{1}{t_0-t_{\rm B}(r)}\int_0^1\frac{dy}{y\sqrt{\Omega_{\rm M}(r)y^{-3}+(1-\Omega_{\rm M}(r))y^{-2}}},
\end{eqnarray}
where $y=R/R_0$.

The CR model has the uniform big-bang time $t_{\rm B}(r)=0$, and the gauge condition is chosen so that 
$R(t_0,r)=r$. In the uniform big-bang model, one functional degree of freedom to specify the model remains. 
In the CR model, this degree of freedom is fixed so that the density-parameter function is given by
\begin{eqnarray}
 \Omega_{\rm M}(r)=\Omega_{\rm M}^{(\rm out)}-(\Omega_{\rm M}^{(\rm out)}-\Omega_{\rm M}^{(\rm in)})e^{-r^2/(2\sigma^2)},
 \label{CR1}
\end{eqnarray}
where $\Omega_{\rm M}^{(\rm out)}=0.7$, $\Omega_{\rm M}^{(\rm in)}=0.242$ and $\sigma =6{\rm Gpc}$.
The Hubble constant at the center is $H_0\equiv H_{\bot 0}(r=0)=74{\rm kms^{-1}Mpc^{-1}}$.

\section{transfer function on the CR model}\label{appendixB}

We use the fitting formula for the matter transfer function $T(k)$ given 
by Eisenstein and Hu~\cite{Eisenstein:1997ik}.
The fitting function defined by equation (16) in Ref.~\cite{Eisenstein:1997ik} 
is determined by the four parameters as
\begin{eqnarray}
 T(k)=T(k;\Omega_{\rm b},\Omega_{\rm c},h,\Theta_{2.7}),
\end{eqnarray}
where $\Omega_{\rm b}$ and $\Omega_{\rm c}$ are 
the cosmological density parameter of baryons and cold dark matter,
$h$ is defined as $h=H_0/(100{\rm kms^{-1}Mpc^{-1}})$
and the CMB temperature is written as $2.7\Theta_{2.7} K$.

\begin{figure}[htbp]
 \begin{center}
 \includegraphics[width=10cm,clip]{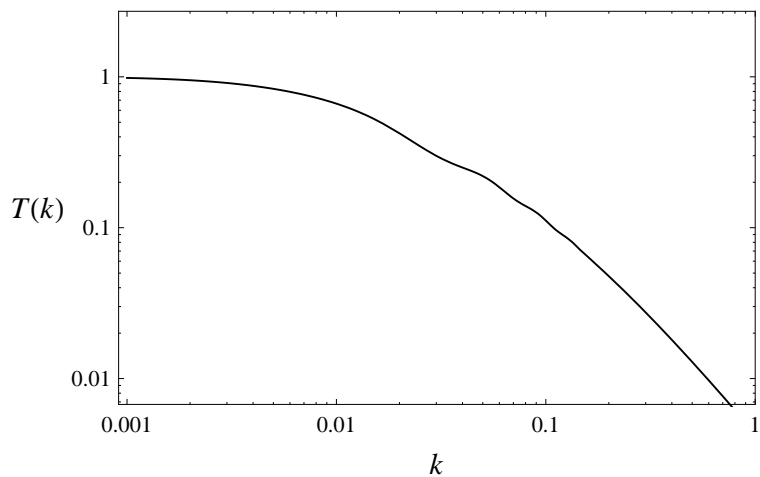}
 \end{center}
 \caption{
Transfer function as a function of $k$
for the FLRW model
with $\Omega_b=0.042, \Omega_c=0.20, h=0.74, \Theta_{2.7}=1.0$.
}
 \label{transfer}
\end{figure}

In this paper, we adopt the following values for these parameters,  
\begin{eqnarray}
 \Omega_b=0.042 ,\quad \Omega_c=0.20,\quad h=0.74,\quad \Theta_{2.7}=1.0\; ,
 \label{paramet}
\end{eqnarray}
where $\Omega_b$, $\Omega_c$ and $h$ are chosen to be the same at the center of the CR model
and $\Theta_{2.7}=1.0$ is assumed.  
This transfer function is plotted as a function of $k$ in Fig.~\ref{transfer}.


\end{document}